\newcommand{\R}{\mathds R}

\documentclass[twoside,a4paper,12pt]{amsart}
\usepackage[active]{srcltx}
\usepackage{times}
\usepackage{dsfont}
\usepackage[bookmarksnumbered]{hyperref}

\renewcommand{\contentsline}[3]{\csname new#1\endcsname{#2}{#3}}
\newcommand{\newchapter}[2]{\bigskip\hbox to \hsize{\vbox{\advance\hsize by -.5cm\baselineskip=12pt\parfillskip=0pt\leftskip=2cm\noindent\hskip -2cm #1\leaders\hbox{.}\hfil\hfil\par}$\,$#2\hfil}}
\newcommand{\newsection}[2]{\medskip\hbox to \hsize{\vbox{\advance\hsize by -.5cm\baselineskip=12pt\parfillskip=0pt\leftskip=2.5cm\noindent\hskip -2cm #1\leaders\hbox{.}\hfil\hfil\par}$\,$#2\hfil}}
\newcommand{\newsubsection}[2]{\medskip\hbox to \hsize{\vbox{\advance\hsize by -.5cm\baselineskip=12pt\parfillskip=0pt\leftskip=3.5cm\noindent\hskip -2cm #1\leaders\hbox{.}\hfil\hfil\par}$\,$#2\hfil}}

\title{Qualitative analysis of collapsing isotropic fluid spacetimes}
\author[R.\ Giamb\`o , \ G.\ Magli]{Roberto Giamb\`o \and Giulio Magli}
\address{\begin{tabular}{lll}
Scuola di Scienze e Tecnologie & & Dipartimento di Matematica\\
Universit\`a di Camerino & & Politecnico di Milano\\
Italy & & Italy\\
\texttt{roberto.giambo@unicam.it} & & \texttt{giulio.magli@polimi.it} \\
\end{tabular}
}

\begin{document}


\theoremstyle{plain}\newtheorem{teo}{Theorem}[section]
\theoremstyle{plain}\newtheorem{prop}[teo]{Proposition}
\theoremstyle{plain}\newtheorem{lem}[teo]{Lemma}
\theoremstyle{plain}\newtheorem{cor}[teo]{Corollary}
\theoremstyle{definition}\newtheorem{defin}[teo]{Definition}
\theoremstyle{definition}\newtheorem{assum}[teo]{Assumption}
\theoremstyle{remark}\newtheorem{rem}[teo]{Remark}
\theoremstyle{definition}\newtheorem{example}[teo]{Example}
\theoremstyle{remark}\newtheorem{step}{\bf Step}
\theoremstyle{plain}\newtheorem*{teon}{Theorem}


\begin{abstract}

{The structure of the Einstein field equations describing the gravitational collapse of spherically symmetric isotropic fluids is analyzed here for general equations of state. A suitable system of coordinates is constructed which allows us, under a
hypothesis of Taylor-expandability with respect to one of the coordinates, to approach the problem of the nature of the final state without knowing explicitely the metric. The method is applied to investigate the singularities of linear barotropic perfect fluids solutions and to a family of accelerating fluids.}
\end{abstract}

\maketitle

\section{Introduction}

Spherical symmetry is a useful test-bed for open problems of astrophysical interest in General Relativity. Among them, a very relevant one is that of the final state of gravitational collapse and, therefore, of the validity of a "Cosmic Censorship"
hypothesis. In particular, the study of spherically symmetric perfect fluid spacetimes is a recurrent topic in relativistic literature, and the final state of the gravitational collapse of spherically symmetric isotropic fluids is still a matter of debate. What makes isotropic fluids' collapse one of the most intriguing problem in gravitational collapse theory is that, on one side, these fluids are useful to model stars in astrophysics and, on the other, that they are a obvious, physically natural  generalization of the so-called Lemaitre--Tolman--Bondi (LTB) dust solutions. LTB models play a distinct role in cosmology as perturbed Friedmann models; in such a framework the effects of pressure become relevant when the models are extended back in time, and inhomogeneous perturbations of the radiation-dominated universe are considered \cite{kra-rev,jsk}.
From the analytical point of view, the LTB models are one of the few known-in-details families of solutions dynamically collapsing to a singularity, and are long known to form both black holes and naked singularities in dependance from the choice of the initial data (the first example was discovered in \cite{c}, and the complete analysis is due to \cite{js}). A class of anisotropic perturbations, whose causal behavior strongly resembles that of LTB models has been studied in \cite{ns}. On the other side, the situation for isotropic fluids is quite less clear; some results are actually known from numerical relativity, in particular for barotropic perfect fluids with linear equation of state: in the spirit of a pioneering work by Choptuik \cite{chop1} on the gravitational collapse of a scalar field  -- yet today one of the cornerstones in this line of research -- these analyses were basically focused to study a one--parameter family of solutions,  aiming to detect a critical value of the parameter separating a branch of solutions with energy dispersion from another branch leading to a black hole. The solution related to that critical parameter is usually a naked singularity (see \cite{lr} and references therein). Naked singularities also occur in self-similar models, as shown by Ori and Piran \cite{op} and Harada \cite{har}. Choptuik himself has worked with Neilsen \cite{nc} to the "ultrarelativistic" case of pressure close to energy density, as well as Snajdr \cite{sn}. Other contributions on the subject were \cite{ec,mais}. Unfortunately, outside the realm of numerical relativity, little is known about the geometry of these spacetimes: whether a singularity is developed, and if that is the case, what is the causal structure of the solution.

The main difficulty in approaching the perfect fluid collapse is, of course, that few exact solutions are known \cite{Kramer, Krasinski}.  So motivated, we developed a conditioned approach, which allows the analisys of perfect fluid spacetimes for general equations of state provided that certain regularity assumptions are satisfied by the (generally unknown) solution. These assumptions essentially require Taylor-expandability of the solution in a special system of coordinates and allow for a quite general picture of barotropic  perfect fluids (with pressure proportional to energy density) as well as for some other cases of interest.
The qualitative picture emerging from these models is quite different from the LTB case. In particular, a crucial role is played by the pressure in the neighborhood of the singular boundary in order to determine the causal structure of the spacetime, as already hinted at in \cite{prepr}, where homogeneous dust collapse is perturbed adding a small amount of pressure.

\section{Isotropic fluids}

Let us consider the general spherical line element in comoving coordinates
\begin{equation}\label{eq:1}
\mathrm ds^2=-e^{2\nu(t,r)}\mathrm dt^2+e^{2\lambda(t,r)}\mathrm dr^2+R(t,r)^2\mathrm d\Omega^2
\end{equation}
Gravity is coupled with a perfect fluid matter tensor of the form
\begin{equation}\label{eq:2}
 T^\mu_\nu=\frac1{8\pi}\text{diag}\{-\epsilon(t,r),p(t,r),p(t,r),p(t,r)\}.\end{equation}
Here $\nu,\,\lambda$ and $R$ are all functions of $(t,r)$ only, such as the energy density $\epsilon$ and the pressure $p$. Denoting by $\dot f$ and $f'$ the partial derivatives with respect to $t$ and $r$ respectively, and introducing the so called Misner--Sharp mass function
\begin{equation}
m=\frac R2(1-(R'e^{-\lambda})^2+(\dot Re^{-\nu})^2),\label{eq:3}
\end{equation}
a complete set of Einstein  equations is given by
\begin{subequations}
\begin{align}
&2m'=\epsilon R^2 R',\label{eq:4a}\\
&2\dot m=-pR^2\dot R,\label{eq:4b}\\
&\dot R'=\dot\lambda R'+\nu'\dot R,\label{eq:4c}\\
&p'=-(\epsilon+p)\nu',\label{eq:4d}
\end{align}
\end{subequations}
As  a consequence of the above equations, the equation of motion
\begin{equation*}
\dot\epsilon=-\left(\frac{2\dot R}R+\dot\lambda\right)(\epsilon+p)
\end{equation*}
also holds.

For obvious physical reasons, the above system is underdetermined until a further physical condition - an equation of state or a kinematic condition on the fluid's motion - has been imposed. The most natural choice to close the system is that of the so-called barotropic equation of state:
\begin{equation}\label{eq:5}
p=p(\epsilon)
\end{equation}
Once the equation \eqref{eq:5} has been prescribed, initial data for the system may be given assigning $R$ at initial time. Without loss of generality we will take
\begin{subequations}
\begin{equation}\label{eq:6a}
R(0,r)=r
\end{equation}
in such a way that an independent set of initial data, for example, is
\begin{align}
&m(0,r)=\mu_0(r),\label{eq:6b}\\
&\dot R(0,r)=\zeta(r),\label{eq:6c}
\end{align}
\end{subequations}

\begin{rem}\label{rem:data}
It is easy to see that all the other initial data are determined by \eqref{eq:6a}--\eqref{eq:6c}. Indeed, we get $\epsilon(0,r)$ from \eqref{eq:4a}  and hence $p(0,r)$ from \eqref{eq:5} and $\dot m(0,r)$ from \eqref{eq:4b}. From \eqref{eq:4d} we obtain $\nu'(0,r)$ which can be integrated to get $\nu(0,r)$ since $\nu(0,0)$ can be set to zero up to time rescaling. Equation \eqref{eq:3} and \eqref{eq:4c} give $\lambda(0,r)$ and $\dot\lambda(0,r)$ respectively. Deriving \eqref{eq:4a} with respect to time we get $\dot\epsilon(0,r)$ and $\dot p(0,r)$ again from \eqref{eq:5}, which can be used in \eqref{eq:4d} derived with respect to $t$ to get $\dot\nu(0,r)$.
\end{rem}

In the following we will be interested in the study of solutions to equations \eqref{eq:4a}--\eqref{eq:4c} modeling a star undergoing complete collapse until a singularity is possibly developed. To describe this situation correctly, a number of physical reasonability conditions must be satisfied.
First of all, the \textit{dominant energy condition} \textit{(DEC)} must hold:
\begin{equation}\label{eq:6}
\epsilon\ge 0,\qquad |p|\le \epsilon.
\end{equation}
Then, we require that there exists an initial time (say, $t=0$) at which the solution is regular, so that the singularities
will be a sincere outcome of the collapse. In particular, the energy density must be finite and outwards decreasing:
\begin{equation}\label{eq:7d}
\lim_{r\to 0^+}\epsilon(0,r)\in\R,\quad\frac{\mathrm d}{\mathrm dr}\epsilon(0,r)\le 0.
\end{equation}
This implies, using \eqref{eq:4a} and \eqref{eq:6a}, that the mass function $\mu_0(r)$ at initial time (see \eqref{eq:6b}) can be chosen to be a regular function at $r=0$ such that
\begin{equation}\label{eq:7aa}
\mu_0(r)=\frac12 m_0r^3+o(r^3).
\end{equation}
Some conditions on the metric must also be imposed, to prevent a bad behavior of the center of symmetry due to the coordinate choice,
so that the polar 'singularity' $r=0$ can be removed using a local Cartesian frame \cite{ref20}:
\begin{equation}\label{eq:8}
R(t,0)=0,\qquad R'(t,0)=e^\lambda(t,0).
\end{equation}
Finally, we also ask for the solution to be free from \textit{shell--crossing singularities}. A sufficient condition, that we will required,  is given by
\begin{equation}\label{eq:9a}
R(t,r)>0\Rightarrow R'(t,r)>0,
\end{equation}
that ensures that shell crossing singularities will not appear prior to shell-focussing singularities,
namely those due to the vanishing of $R(t,r)$. Shell--crossing singularities usually correspond to Tipler-weak divergences of the curvature, though in some cases the spacetime extension problem beyond them has been discussed \cite{nolan-sc}.
It must also be observed that \eqref{eq:9a} is only sufficient, since in principle $R'$ may vanish in such a way that $m'/R'$ remains finite, see \eqref{eq:4a}, and then no shell--crossing singularity actually takes place \cite{joshi-sc}.

To model collapse dynamics, we shall always require $\dot R\le 0$, which results in choosing the negative sign from extracting the square root from \eqref{eq:3}:
\begin{equation}\label{eq:7}
\dot R=-e^{-\nu}\sqrt{\frac{2m}{R}-1+(R'e^{-\lambda})^2}.
\end{equation}
In order to obtain the model of a colapsing compact object, a matching with an external space should be performed at a boundary surface, which  in comoving coordinates can be always taken as $\{r=r_b=\rm const\}$. Using  Israel--Darmois junction condition, from \eqref{eq:4b} it follows easily that a necessary and sufficient condition to match the solution with a Schwarzschild exterior is that the pressure $p$ vanish on the matching surface.
A milder condition which also fulfills physical requirements is to ask for a matching to a radiating metric such as the generalized Vaidya spacetime, as done for instance in \cite{cs}.

\section{The qualitative analysis}\label{sec:qual}
To approach the problem of the singularities without knowing explicitely the metric it is convenient to use a system of coordinates where the singularity curve is mapped to a "point". A relatively suitable system \cite{ns,uw} is that of the area radius coordinates $(r,R,\theta,\phi)$ where the shell focussing singularity is mapped into $R=0$. This proves to be a good choice for the study of charged dust \cite{ori22} and also for vanishing radial stress models \cite{ref11,ref20}. However, in view of \eqref{eq:10}, this choice has the disadvantage of mapping the center of symmetry of the star ($r=0$) in the point ($R=r=0$) regardless of whether the center is singular or not. This suggests a slight variant of the above choice, first used in \cite{PJM}: coordinates $(r,a,\theta,\phi)$ where
$$
a=\frac{R(t,r)}{r}.
$$
In this way the Cauchy surface of initial data $t=0$ corresponds to the set $a=1$, while $a=0$ is the singularity curve. The price is that one has to be careful because it is not garanteed that all the values in the interval $\{0<a<1\}$ are
dynamically admissible. This can be inferred from the study of the sign of the quantity under square root in \eqref{eq:7}, which now is a function of $(r,a)$. The region where this quantity becomes negative cannot, of course, be reached. Moreover, also when the whole strip $\{0<a<1\}$ is allowed, it must always be controlled that the singularity is developed in a finite amount of comoving time.

To write the system of Einstein equations in this new setting we will introduce the following functions of $(r,a)$:
\begin{equation}\label{eq:9}
\gamma=\frac p\epsilon,\quad\,B=e^\lambda,\quad\,F=e^\nu,\quad\,Y=R'e^{-\lambda}.
\end{equation}
Moreover, recalling \eqref{eq:7aa}, we will use the function
\begin{equation}\label{eq:10}
M(r,a)=\frac{2m}{r^3}
\end{equation}
in place of Misner--Sharp mass \eqref{eq:3}. Finally, we also make the positions
\begin{equation}\label{eq:11}
w=a',\qquad z=\dot a,
\end{equation}
that will be used to trace back information on the comoving system from the current framework. In this way equations \eqref{eq:4a}--\eqref{eq:4b} become
\begin{subequations}
\begin{align}
&3M+M_rr+wrM_a-\epsilon a^2(wr+a)=0,\label{eq:12a}\\
&M_a+\gamma\epsilon a^2=0,\label{eq:12b}
\end{align}
(where subscripts $a$ and $r$ stand for partial derivatives) whereas equations \eqref{eq:4c}--\eqref{eq:4d} take the form
\begin{align}
&(k+1)\epsilon(wr+a)Y_a+Y\left[(k\epsilon)_r+w(k\epsilon)_a\right]r=0,\label{eq:12c}\\
&r(F_r+wF_a)Y-(wr+a)Y_aF=0,\label{eq:12d}
\end{align}
where $z$ and $w$ are given by
\begin{align}
z&=-F\left(\frac Ma+\frac{Y^2-1}{r^2}\right)^{1/2},\label{eq:12e}\\
w&=\frac{YB-a}r.\label{eq:12f}
\end{align}
Finally, an equation is needed to express the compatibility property between $w$ and $z$ introduced in \eqref{eq:11}, i.e. $\dot w=z'$, that in the $(r,a)$ system reads as
\begin{equation}\label{eq:12g}
z_r+wz_a-zw_a=0.
\end{equation}
\end{subequations}

\begin{rem}\label{rem:data2}
Equations \eqref{eq:12a}--\eqref{eq:12g} is a system of  PDE in the unknown $B,\,F,\,Y,\,\epsilon,\,M$ under the functional dependencies \eqref{eq:12e}--\eqref{eq:12f}, and can be closed as before with the prescription of an equation of state
\begin{equation}\label{eq:13}
\gamma=\gamma(\epsilon).
\end{equation}
A  set of independent initial data at $a=1$ can be proved - similarly to Remark \ref{rem:data} - to be $M(r,1)$ and $B(r,1)$, whereas condition \eqref{eq:6a}, using \eqref{eq:12f}, states that
\begin{equation}\label{eq:14}
w(r,1)=0.
\end{equation}
\end{rem}

\begin{rem}\label{rem:wz}
Equation \eqref{eq:6} states that the 1-form
$$
\frac1z\left(\mathrm da-w\mathrm dr\right)
$$
is an exact differential. The comoving time $t(r,a)$ is its integral, thus given by
\begin{equation}\label{eq:t1}
t(r,a)=t(0,1)-\int_0^r\frac{w(s,1)}{z(s,1)}\,\mathrm ds+\int_1^a\frac1{z(r,\bar a)}\mathrm d\bar a.
\end{equation}
and so, chosing $t(0,1)=0$ and using \eqref{eq:14} we get
\begin{equation}\label{eq:t2}
t(r,a)=-\int_a^1\frac1{z(r,\bar a)}\mathrm d\bar a.
\end{equation}
\end{rem}

\begin{example}\label{ex:dust}
As a particular case here we briefly recall the well known dust model,  a trivial isotropic model since pressure vanishes ($p=0$). Equation \eqref{eq:4b} implies $m=m(r)$ and
in comoving coordinates the metric is given by
$$
-\mathrm dt^2+\frac{R'^2}{1+f(r)}\,\mathrm dr^2+R^2\,\mathrm d\Omega^2,
$$
where
$$
\dot R=-\sqrt{\frac{2m(r)}{R}+f(r)},
$$
and $f(r)$ is a free initial data function which is equivalent to $\zeta(r)$ (see \eqref{eq:6c}). Let us consider for the sake of simplicity the so--called marginally bound case, corresponding to the choice $f(r)\equiv 0$. Then $R'=e^{\lambda}$ and $\dot R=-\sqrt{\frac{2m(r)}{R}}$, from which
$$
R(t,r)=r(1-k(r) t)^{2/3},\qquad\text{with\ }k(r)=\frac32\sqrt{\frac{2m(r)}{r^3}}.
$$
Let us rewrite this model using $(r,a)$ coordinates. We have $F(r,a)=Y(r,a)=1$, and $M=\tfrac49k(r)^2$. Consequently,
\begin{align*}
&w(r,a)=-\frac{2k'(r)}{3k(r)}\left(\frac1{\sqrt a}-a\right),\\ &z(r,a)=-\frac{2 k(r)}{3\sqrt a},\\
&B(r,a)=a\left(1-\frac{2rk'(r)}{3k(r)}\left(\frac1{a\sqrt a}-1\right)\right),
\end{align*}
and $\epsilon(r,a)$ can be obtained from \eqref{eq:12a}.
\end{example}

Notice that an interesting property of the dust solutions above, which is actually the reason that allowed previous studies to gain a complete picture of the nature of their singularities, is that all the relevant quantities
can be expanded in power series with respect to $r$ around $r=0$ if so does the initial datum $k(r)$ (which of course is non-vanishing in $r=0$). This property inspires the work we are carrying out here, since we are now going to {\it assume}
this behaviour on the {\it general} solution to equations \eqref{eq:12a}--\eqref{eq:12g}. Accordingly, we set:

\begin{defin}\label{as:exp}
A solution is said to be {\it r-expandable} if $B,\,F,\,Y,\,\epsilon$ and $M$ are
regular up to $r=0$ with respect to the variable $r$. In other words, for $n$ sufficiently large, each of them can be
written in the form
$$
G(r,a)=\sum_{i=0}^n G_i(a) r^i+o(r^n),\,\forall r\in [0,r_b],\,\forall a\in]0,1]
$$
where $o(r)$ above and hereafter must clearly be intended as a function of both $(r,a)$.
\end{defin}

For such solutions the model equations can be expanded with respect to $r$, in order to obtain relations between the Taylor coefficients. Some of these relations are fixed directly by the regularity conditions.
First of all, local flatness means that $R'e^{-\lambda}\to 1$ (that is $B\to 1$) approaching $r=0$, $a>0$. Moreover, up to time reparameterization, we can suppose $\nu\to 1$ as $r\to 0, a>0$. All these facts generate conditions $F_0(a)=B_0(a)=1$. Now, the integral in \eqref{eq:t2} evaluated in $a=0$ reads
$$
\int_0^1 \sqrt{\frac{r}{2 Y_1(\bar a)}} +o(r)\,\mathrm d\bar a.
$$
If this integral exists and is finite, then its limit as $r\to 0$ must be zero, but this in turn would mean that the center is already singular at the initial comoving time. Thus, physical reasonability demands
\begin{equation}\label{eq:Y1}
Y_1(a)=0.
\end{equation}

It must also be remarked that under the above assumptions the function $w(r,a)$ can be developed with respect to $r$ and one has
$$
w(r,v)=\frac{B _0(a)-a}{r}+B_1(a)+o(1).
$$
Actually $w$ is not one of the unknown functions for which we require Taylor expandability.  However, one would like to have $w(r,a)$ continuous up to the regular center $r=0$: indeed, we already know (see \eqref{eq:14}) that $w(r,1)=0$, in order to be able to integrate the 1-form $\mathrm dt$ and calculate $t(r,a)$ along the path suggested in integral \eqref{eq:t1}. But in principle, one should be able to perform integration along the path $\{(0,\bar a)\,:\,v\le \bar a\le 1\}\cup\{(s,a)\,:\,0\le s \le r\}$, because otherwise the time $t=0$ would be a sort of ``privileged'' time for the central shell, which is manifestly unphysical. For this reason we will consider the quite natural cases where $w(r,a)$ is continuous up to the regular center, and then $B_0(a)=a$. This also implies, developing \eqref{eq:2} and
\eqref{eq:5} in $r$ up to order 1 and 0 respectively, we get $F_1(a)=0$ and
\begin{equation}\label{eq:e0}
\epsilon_0(a)=\frac{3 M_0(a)}{a^3}.
\end{equation}

Under the above assumptions, the field equations read as follows:
\begin{align}
&\left(4 M_1(a)-a^3 \epsilon _1(a)+B _1(a) \left(-\frac{3 M_0(a)}a+ M_0'(a)\right)\right) r+o(r)=0,\label{eq:2-1}\\
&\frac{-aM_1'(a)+B _1(a) \left(2 M_0'(a)-a M_0''(a)\right)} {a^3}\,r+o(r)=0,\label{eq:4-1}\\
&\left(2 F _2(a)-a Y_2'(a)\right) r^2+o(r^2)=0,\label{eq:5-1}\\
&\frac12a^{-2}\left({2 Y_2(a)+\frac{M_0(a)}{a}}\right)^{-\tfrac12}\left[-a\left(M_1(a)+2 aY_3(a) -2(M_0(a)+2aY_2(a))B_1'(a)\right) \right.\label{eq:6-0}\\
&\qquad\left.+B_1(a)\left(M_0(a)-a(M_0'(a)+2aY_2'(a))\right)\right]+o(1)=0.\nonumber
\end{align}
One could at this point let the $M_i$'s free, together with $Y_2$. On the other hand, as \eqref{eq:4-1} suggest, finding $B_1$ from $M_0$ and $M_1$ is possible only under the condition that
\begin{equation}\label{eq:special}
\left(2 M_0'(a)-a M_0''(a)\right)\ne 0.
\end{equation}
As a consequence, in order not to lose generality, we prefer to let $B_1(a)$ free and find the expression for $M_1(a)$:
\begin{equation}\label{eq:M1}
M_1(a)=m_1+\int_a^1\tau^{-1}B_1(\tau)\left(\tau M_0''(\tau)-2M_0'(\tau)\right)\,\mathrm d\tau,
\end{equation}
Then, through equation \eqref{eq:2-1} we get $\epsilon_1(a)$, equation \eqref{eq:5-1} yields $F_2(a)$, and \eqref{eq:6-0} gives $Y_3(a)$. This scheme can be iterated, in such a way that:
\begin{enumerate}
    \item\label{itm:1} first, the leading term of \eqref{eq:12a} at $n$th order gives the relation $M_n'(a)=f_n$, where $f_n$ is a regular function depending on $M_0,Y_2,B_1,\ldots,B_{n}$  and possibly their derivatives, and then can be integrated to find $M_n(a)$;
    \item the leading term of \eqref{eq:2} at $n$th order gives an algebraic relation $\epsilon_n(a)=g_n$ algebraically, where $g_n$ is a regular function depending on $M_0,Y_2, B_1,\ldots,B_{n}$ and possibly their derivatives;
    \item same as above, \eqref{eq:5} leading term at $(n+1)$th order gives $F_{n+1}(a)$ as functionally dependent on $M_0,Y_2, B_1,\ldots,B_{n}$ and possibly their derivatives, through an algebraic relation;
    \item finally, \eqref{eq:6} at $(n-1)$th order gives  the functional dependence of $Y_{n+2}(a)$ in terms of $M_0,Y_2$,  $B_1,\ldots,B_{n}$ and possibly their derivatives.
\end{enumerate}

With the above iterative scheme, the coefficients of the solutions can be determined up the to freedom in choosing $M_0(a)$, $Y_2(a)$, $B_i(a)$ and $M_i(1)$ (with $i\ge 1$). $M_i(1)$ is the initial condition coming from step
\eqref{itm:1} above which is the only one involving the integration of a differential equation. Of course, specifying also the equation of state \eqref{eq:13} allows us to choose $M_0$  and all coefficients $B_i(a)$ (up to their initial data $B_i(1)$)
in such a way that the only  freedom left is in the choice of the function $Y_2(a)$.
It is this function that encodes all the degrees of freedom which are left, pertaining to the initial data and to the matching with an external solution or the imposition of asymptotic behavior leading to local flatness at space infinity. Indeed, as we will see in the examples below, $M(r,1)$ and $B(r,1)$ are given in terms of $Y_2(a)$ and all its derivatives evaluated in $a=1$. Interestingly enough, the initial data in \eqref{eq:14} impose a constraint that results in the vanishing of all odd order coefficients; the matching conditions with generalized Vaydia spacetime instead do not add constraints, since the resulting mass of the exterior solution is fixed by the internal one. We stress however that, of course, the method does not guarantee a priori convergence of the series, that would require a--priori estimates on the remainder.

\begin{example}\label{ex:sf}
The dust models recalled in \ref{ex:dust} are of course the first example of a class fulfilling the above assumptions. A second relevant example is that of  shearfree perfect fluids (see e.g. \cite{Kramer,Krasinski,bmj}). With the notations used here the shearfree conditions can be written as
\begin{equation*}
B(r,a)=h(r) a
\end{equation*}
with $h(r)$ an arbitrary function. For the sake of simplicity we consider here only the (homogeneous) case $h(r)=1$. In this way all the arbitrary functions $B_i(a)$ are set to zero for $i\ge 1$, the initial conditions are constrained to satisfy $M_i(1)=Y_2(1)=0$ ($i\ge 1$), and this completely sets all arbitrary functions except $M_0(a)$ and $Y_2(a)$. Their choice determines the leading term of energy and pressure near the centre, indeed:
\begin{align*}
\epsilon(r,a)&=\frac{3 M_0(a)}{a^3}-\frac{3 r^2 Y_2(a) \left(a M_0'(a)-3 M_0(a)\right)}{2 a^3}+o(r^2),\\
p(r,a)&=-\frac{M_0'(a)}{a^2}
+\frac{r^2 \left(Y_2(a) \left(a M_0''(a)-2 M_0'(a)\right)+\left(a M_0'(a)-3 M_0(a)\right) Y_2'(a)\right)}{2 a^2}+o(r^2).
\end{align*}

\end{example}

\section{The nature of the singularities}

In the present section we study the formation and nature of singularities for some physically interesting models of isotropic fluids under the hypothesis of $r$-expandability. In particular, we want to investigate the correlation between models which may generically give rise to naked singularities and the behavior of the pressure in the late stage of the collapse, starting from a situation where this quantity diverges together with the energy density.

The singularity forms only if the (comoving) time of collapse is finite. Recalling \eqref{eq:Y1}, the function $t(r,a)$ \eqref{eq:t2} becomes
$$
t(r,a)=\int_0^1 \frac{1}{\sqrt{2 Y_2(\bar a)+\frac{M_0(\bar a)}{\bar a}}}+ o(1)\,\mathrm da
$$
and then supposing that \eqref{eq:t2} exists finite, the time of collapse of the central shell is given by
\begin{equation}\label{eq:ts0}
t_s(0):=\int_0^1 \frac{1}{\sqrt{2 Y_2(\bar a)+\frac{M_0(\bar a)}{\bar a}}}\,\mathrm da.
\end{equation}
In all the examples that we are going to study, we will be concerned with those collapsing models where the free function $Y_2(a)$ is  regular up to $a=0$. We stress that this condition, although very reasonable, does not include all the physically relevant cases, as the time of collapse can of course be finite also with a diverging behavior of $Y_2(a)$. The analysys of such cases is deferred to a future work.

To study the behavior of the central singularity, we will use a method already successfully exploited for other models \cite{Nolan, ns, uw}. The method consists in investigating the existence of radial null geodesic by studying the properties of the differential equation satisfied by these geodesics, which reads
\begin{equation}\label{eq:geo}
\frac{\mathrm da}{\mathrm dr}=\Phi(r,a)=:z(r,a) \frac{B(r,a)}{F(r,a)}+w(r,a).
\end{equation}
Of course, the right hand side is not defined at $r=a=0$ and then standard ODE theory does not apply. However,
a remarkable property can be proved that involves the {\it apparent horizon} curve $a_h(r)$. The apparent horizon is the boundary of the region of trapped surfaces; in spherical symmetry it is the curve implicitly defined by the equation $R=2m$ (see for instance \cite{gc}).
It can be proved that $a_h(r)$ is a
is a supersolution of \eqref{eq:geo} -- i.e. $\tfrac{\mathrm da_h}{\mathrm dr}\ge\Phi(r,a_h(r))$.
Then if a subsolution $a_*(r)$ exists such that $a_*(0)=0$ and $a_*(r)>a_h(r)$ for $r>0$, comparison arguments in ODE theory ensure the existence of infinite light rays emerging from the central singularity and "living" in the untrapped region.
We refer the reader to \cite[Theorem 2.5]{uw} for more details about the use of supersolutions and subsolutions to find solutions to \eqref{eq:geo}.

\begin{rem}\label{rem:gloc}
In principle, one may argue that the singularities emerging from this approach are only locally naked. As a matter of fact, however, prolongation of the metrics in such a way that the naked singularity is visible to far-away observers is usually possible \cite{gloc,gc}.
\end{rem}

\subsection{Linear equations of state} \label{sec:lin}
The first model we consider is that of a linear pressure--density relationship $p(\epsilon)=\beta\epsilon$, where the constant $\beta\in[-1,1]$ to comply with the DEC \eqref{eq:6}. The dominant energy condition thus allows for the cases $\beta>0$ - "standard" barotropic fluids - but also solutions with negative pressures (tensions) up to the model generating anti-de Sitter space, for which $\beta=-1$.

The equation of state fixes
\begin{equation}\label{eq:M0}
M_0(a)=m_0 a^{-3\beta},
\end{equation}
Moreover, we have for any $n\ge 1$ a condition expressing the vanishing of the $n$th order coefficient of $k(r,v)$. If $\beta$ is not zero ($\beta=0$ corresponds to a dust) this results in fixing completely all the $B_n$'s up to the initial data, that are completely determined by the only function left to be chosen (i.e., $Y_2(a)$). Indeed, $M(r,1)$ and $B(r,1)$ (which form a set of independent data for this problem, see Remark \ref{rem:data2}) are given by
\begin{multline*}
M(r,1)=m_0-\frac{3 \left((1+\beta ) m_0 Y_2'(1)\right) r^2}{10 \beta }\\
-\frac{3}{560 \beta ^2} \left((1+\beta ) m_0 \left(m_0 \left(9 \beta ^2 Y_2'(1)+7 Y_2''(1)-3 \beta  \left(5 Y_2'(1)+3 Y_2''(1)\right)+2 Y_2{}^{(3)}(1)\right)\right.\right.\\
\left.\left.+2 \left(Y_2'(1) \left(-(4+13 \beta ) Y_2'(1)+Y_2''(1)\right)+\right.\right.\right.\\
\left.\left.\left.Y_2(1) \left((4-22 \beta ) Y_2'(1)+(8-6 \beta ) Y_2''(1)+2 Y_2{}^{(3)}(1)\right)\right)\right)\right) r^4+o(r^4),
\end{multline*}
and
\begin{multline*}
B(r,1)=1-Y_2(1) r^2+\frac{r^4}{20\beta}\cdot\\
{\left[m_0 \left((-5+3 \beta ) Y_2'(1)-2 Y_2''(1)\right)+2 Y_2(1) \left(3 \beta  \left(5 Y_2(1)+2 Y_2'(1)\right)-2 \left(Y_2'(1)+Y_2''(1)\right)\right)\right]}\\+o(r^4).
\end{multline*}
Therefore, as it can be seen, $Y_2(1)$ determines $B_2(1)$,
$Y_2'(1)$ determines $M_2(1)$, $Y_2''(1)$ determines $B_4(1)$,
$Y_2{}^{(3)}(1)$ determines $M_4(1)$ and so on.
Noticeably enough, the expansion of these terms are forced to contain only only even--power terms.
Using these expansions, it is possible to study gravitational collapse in a neighborhood of the centre. In fact,  we can calculate the expression of the apparent horizon curve $a_h(r)$ as follows:
$$
a_h(r)=m_0^{\tfrac1{1+3\beta}} r^{\tfrac{2}{1+3\beta}}+o(r^{\tfrac{2}{1+3\beta}}),
$$
when $\beta\ne-1/3$ -- but see below.
Clearly, a special role is played by the quantity $\bar\beta:=1+3\beta$.

In fact, if $\bar\beta<0$, there exists a right neighborhood of $r=0$ such that $M(r,a)r^2<a$, $\forall a\in]0,1]$, and then the apparent horizon does not form; this suffices to conclude that the singularity is globally naked (this behaviour of barotropic perfect fluids was already found, under different assumptions, in \cite{coop}). Also the case $\bar\beta=0$ arises as a limit case of the above, since it is found that $M(r,a)r^2-a=-a(1-m_0 r^2+o(r^2))$ and then no horizon forms near the center.

If  $\bar\beta>0$ we must study the null radial geodesic equation. It is sufficient to study the behavior of this equation along test curves of the kind $a_\lambda(r)=(\lambda r^2)^{\tfrac{1}{\bar\beta}}$ (with $\lambda>m_0$ in order that the curve stays above $a_h(r)$). In fact these are the curves that, translated in comoving coordinates, leave $r=0$ together with the apparent horizon. The condition for these curves to be subsolutions of \eqref{eq:geo} reads
\[
-\frac{\lambda ^{\frac{1}{\bar\beta}} \left(\bar\beta r \sqrt{\frac{m_0}{\lambda  r^2}+2 \upsilon(r,\lambda) }+2\right)}{
\bar\beta r}>0,
\]
where $\upsilon$ is a regular function on $r=0$ depending on $Y_2$. Clearly this condition is not satisfied
by any  positive $\lambda$, and therefore the singularities are covered and the solutions form blackholes. We summarize the result in the following

\begin{prop} In the collapse of an isotropic, r-expandable fluid solution with linear equation of state $p=\beta \epsilon$, $\beta\in[-1,1]\setminus\{0\}$, the singularity is naked if $\beta\le-1/3$, while it is covered if $\beta>-1/3$.
\end{prop}

\begin{rem}\label{rem:sec}
Recalling that the \textit{strong energy conditions} (SEC) for isotropic models reads
$$
\epsilon+p\ge 0,\qquad\epsilon+3p\ge 0,
$$
then, remarkably enough, the values of $\beta$ ensuring the SEC - and therefore the "attractive" behavior of gravity  -
also ensure horizon formation, covering the singularity (except of course the dust collapse $\beta=0$ \cite{uw}, and the borderline case $\beta=-1/3$).
\end{rem}

\subsection{Fluids with acceleration vanishing at the singularity}
In the above described example, linear equations of state $p=\beta\epsilon$ with $\beta>-1/3$ with bounded $Y_2(a)$ - which can be seen as "perturbations" of the dust solutions with the same data - always lead to blackhole formation. The presence of pressure drives the final state always to a covered singularity, at least within the hypotheses considered. Clearly,
the pressure diverges at the singularity as well as the density in these models; it is therefore interesting to investigate cases in which the pressure stays finite at the singularity, to check if pressure divergence is necessary to halt naked singularity formation. If the fluid is barotropic this clearly requires "exotic" equations of state, since $\lim_{\epsilon \to \infty} p(\epsilon)$ must remain bounded.

Actually, it must first be remarked that under the assumption made before, the case in which $p$ remains bounded but non-zero
as the fluid collapses does not lead to singularity formation. Indeed, the leading term of the pressure in general is given by $-\frac{M_0'(a)}{a^2}$. Considering a pressure tending to a nonzero constant as $a\to 0$ fixes the asymptotic behavior of $M_0(a)=m_0 a^3$ as $a\to 0$, which fixes the leading behavior of the pressure as follows:
$$
p(r,a)=-3 m_0-\frac{5 m_2  Y_2'(a)}{4 a^2}r^4+o(r^5),
$$
but at the same time the function $z(r,a)$ is given by
$$
z(r,v)=
-{\sqrt{a^2 m_0+2 Y_2(a)}}+o(1).
$$
Then to have $p(r,a)$ not diverging as $a\to 0$, and excluding the non generic case $m_2\ne 0$, the function $Y_2(a)$ must be such that the integral
$\int z^{-1}\,\mathrm da$ does not converge in a right neighborhood of zero, and then the singularity forms in an infinite amount of comoving time, resulting in an eternally collapsing, but regular, spacetime.

Thus we search for models where the pressure vanishes dynamically as the singularity forms.
Now, let us recall that the acceleration of the fluid in comoving coordinates is given by $a_\mu=
\nu'\delta_\mu^r$, therefore, it can be uniquely characterized by
the scalar $A:=\sqrt{a_\mu a^\mu}$.
Since pressure and acceleration are connected by  relation
$$
a_\mu=-\frac{p'}{\epsilon+p}\delta^r_\mu
$$
and the dust (zero pressure) solutions are also non-accelerating solutions, a simple way
to model such situation is to study those isotropic fluids with non-vanishing acceleration, such that acceleration goes to zero in the approach to the singularity. To construct such models we consider the case $M_0(a)=m_0\in\R$ and use
$
\nu'=\frac{R'Y_{,a}}{r Y},
$ which gives
\begin{equation}\label{eq:acc}
A=\frac{Y_{,a}}r =Y_2'(a) r+o(r) \ ,\  p=-\frac{3 m_0 Y_2'(a)}{2 a^2}r^2+o(r^2)
\end{equation}
Thus the pressure will be bounded at the singularity whenever $Y_2(a)$ is constant; consequently we have $
Y(r,a)=1+y_2 r^2+o(r^2)
$
where $y_2$ is a constant as well, and it is also found $M_2(a)=m_2$ and
$$
A=\frac{2 a m_2+4 a^3 y_2 B_2''(a)+m_0 \left(-3 a y_2-2 B_2(a)-a B_2'(a)+2 a^2 B_2''(a)\right)}{4 a^3}r^3+o(r^3),
$$
so it makes sense to suppose that $A$ goes like $\kappa R^3=\kappa a^3 r^3$, with $\kappa\in\R$, determining the behavior of $B_2(a)$ up to a constant $b_2$ -- the other constant is given imposing the condition $R'=1$ at $a=1$.

Since, in this situation, the apparent horizon $a_h(r)$ goes like $m_0 r^2+m_2 r^4+ o(r^4)$, with $m_0>0$, and $m_2<0$ in order to have a outward decreasing energy function at the initial time, then one obtains the condition for a central naked singularity, that is existence of a subsolution of equation \eqref{eq:geo} of the form $a_*(r)=\lambda r^2$ with $\lambda>m_0$. Interestingly enough, the very same condition is that preventing the formation of shell crossing singularities near the centre, as can be  seen with some algebra. Since $y_2\ne 0$ yields a quite complicate expression, here we report only the case when $y_2=0$, that turns out to be
$$
\frac{2 \left(-12 \kappa +39  b_2 m_0+26 m_2\right)}{195 m_0}<0
$$
The above becomes a condition on the coefficients of the functions $M_0, M_2$ and $B_2$, and $\kappa$ as well.  Thus, we conclude that

\begin{prop}\label{thm:chap2}
In the complete collapse of an isotropic $r$-expandable fluid, if acceleration vanishes at the singularity a central naked singularity forms.
\end{prop}

\section{Conclusions}

We are still far from a complete understanding of perfect fluid collapse even in spherical symmetry. However, from the results  discussed above - which can be considered as conditioned results, since we assumed a priori a certain regularity of the solutions - the role of pressure appears clearly. Pressure influences the qualitative behavior of the solution and therefore, the causal structure of the collapsing model. In the linear case $p=\beta\epsilon$ the pressure --
when is nonzero, thus excluding LTB model -- diverges with the energy density in the approach to the singularity, and in the cases implying formation of the horizon, this completely hides the singularity.
These models also contains some interesting cases where on the contrary the horizon does not even form and then the singular boundary is globally naked. This behaviour was already devised in \cite{ihm} where examples showing a central naked singularity of this kind were obtained.

A simpler picture arises when the equation of state is perturbed in such a way that the pressure goes to zero as the energy diverges -- here, these models \textit{are} proper dust perturbations, since LTB solutions are recovered in the limit $\mu\to 0$ -- and here a central naked singularity takes place.

To conclude, boundedness of pressure near the singular boundary appears to be a key ingredient to produce counterexamples to cosmic censorship in the isotropic case since - within the assumptions of expandability used -
finite, non identically zero pressures always lead to a central naked singularity. Isotropy, far from simplifying the geometry of the spacetime, actually adds a series of interesting situations which do not appear in the examples already known of anisotropic spacetimes  (see e.g. \cite{uw} and references therein) where both tangential and radial pressures diverge at the singularity and the endstates are quite similar in structure to those of the dust solutions.
Of course, to get a complete picture one should be able to prove convergence theorems for the  series of the unknown functions of the system. This might in principle cut out some of the examples discussed here.


\end{document}